\documentclass[a4paper,12pt,reqno,english]{amsart}
\usepackage[utf8]{inputenc}
\usepackage[T1]{fontenc}
\usepackage{mathrsfs}
\usepackage{dsfont}
\usepackage{amsmath}
\usepackage{amssymb}
\usepackage{amsthm}
\usepackage{amsfonts}
\usepackage{amstext}
\usepackage{amsopn}
\usepackage{amsxtra}
\usepackage[foot]{amsaddr}
\usepackage{dsfont}
\usepackage{color}
\usepackage{esint}
\usepackage{graphicx}
\usepackage[margin=3cm]{geometry}
\usepackage[iso]{isodate}
\usepackage{xcolor}
\usepackage{todonotes}
\setuptodonotes{inline}
\usepackage[shortlabels]{enumitem}
\usepackage{bbm}
\usepackage{braket}
\usepackage{hyperref}
\usepackage[capitalise,sort]{cleveref}

\usepackage{aliascnt}

\usepackage{lipsum}

\usepackage{chngcntr}
\counterwithin{figure}{section}
\counterwithin{equation}{section}
\counterwithin{table}{section}

\usepackage[style=alphabetic,
  sorting = nyt,
  sortcites=true,
  giveninits=true,
  date=year,
  isbn=false,
  maxbibnames=99,
  maxalphanames=7,
  backend=biber]{biblatex}
\DeclareSourcemap{
  \maps[datatype=bibtex]{
    \map[overwrite]{
      \step[fieldsource=doi, final]
      \step[fieldset=url, null]
      \step[fieldset=eprint, null]
    }  
  }
}

\DeclareSourcemap{
  \maps[datatype=bibtex, overwrite]{
    \map{
      \step[fieldset=language, null]
      \step[fieldset=month, null]
      \step[fieldset=urldate, null]
      \step[fieldset=issue, null]
    }
    \map{
      \pertype{article}
      \step[fieldset=eprint, null]
    }
  }
}

\AtEveryBibitem{%
  \clearlist{language}%
  \clearlist{month}%
  \clearlist{urldate}%
  \clearfield{urldate}%
  \ifentrytype{inproceedings}
    {\clearlist{booktitle}%
     \clearlist{publisher}%
     \clearlist{location}}
}

\renewbibmacro{in:}{}

\DeclareFieldFormat[misc]{title}{\mkbibquote{#1}}
\DeclareFieldFormat[article,inproceedings]{volume}{\mkbibbold{#1}}
\DeclareFieldFormat[inproceedings]{series}{\mkbibitalic{#1}}
\DeclareFieldFormat{titlecase}{\ifcurrentfield{journaltitle}{#1}{\MakeSentenceCase{#1}}}
\renewcommand*{\bibfont}{\footnotesize}
\appto\bibfont{\setlength{\emergencystretch}{2em}}
\usepackage{seqsplit}

\theoremstyle{plain}

\newtheorem{theorem}{Theorem}[section]
\newaliascnt{corollary}{theorem}\aliascntresetthe{corollary}
\newaliascnt{lemma}{theorem}\newtheorem{lemma}[lemma]{Lemma}\aliascntresetthe{lemma}
\newaliascnt{proposition}{theorem}\aliascntresetthe{proposition}

\theoremstyle{definition}

\newaliascnt{notation}{theorem}\newtheorem{notation}[notation]{Notation}\aliascntresetthe{notation}
\newaliascnt{remark}{theorem}\newtheorem{remark}[remark]{Remark}\aliascntresetthe{remark}
\newaliascnt{definition}{theorem}\newtheorem{definition}[definition]{Definition}\aliascntresetthe{definition}
\newaliascnt{assumption}{theorem}\newtheorem{assumption}[assumption]{Assumption}\aliascntresetthe{assumption}

\crefname{assumption}{Assumption}{Assumptions}

\newcommand{\R}{\mathbb{R}}
\newcommand{\Z}{\mathbb{Z}}

\newcommand{\N}{\mathbb{N}}

\newcommand{\ii}{\infty}

\newcommand{\cP}{\mathcal{P}}
\newcommand{\cT}{\mathcal{T}}

\newcommand{\cV}{\mathcal{V}}

\newcommand{\cU}{\mathcal{U}}

\newcommand{\cD}{\mathcal{D}}

\newcommand{\cH}{\mathcal{H}}

\newcommand\pscal[1]{{\ensuremath{\left\langle #1 \right\rangle}}}
\newcommand{\norm}[1]{ \left\| #1 \right\|}
\newcommand{\snorm}[1]{ \| #1 \|}
\newcommand{\abs}[1]{{\left | #1 \right |}}
\newcommand{\nnorm}[1]{{\left\vert\kern-0.25ex\left\vert\kern-0.25ex\left\vert #1 
    \right\vert\kern-0.25ex\right\vert\kern-0.25ex\right\vert}}
\newcommand{\snnorm}[1]{{\vert\kern-0.25ex\vert\kern-0.25ex\vert #1 \vert\kern-0.25ex\vert\kern-0.25ex\vert}}

\let\Re\relax
\let\Im\relax
\DeclareMathOperator{\Im}{Im}
\DeclareMathOperator{\Re}{Re}

\DeclareMathOperator{\Ran}{Ran}

\renewcommand{\geq}{\geqslant}
\renewcommand{\leq}{\leqslant}

\renewcommand{\hat}{\widehat}
\renewcommand{\tilde}{\widetilde}

\def\la{\langle}
\def\ra{\rangle}

\newcommand{\nn}{\nonumber}
\newcommand{\rd}{\mathrm{d}}
\newcommand{\dx}{\rd x}

\title[Sobolev theory for the inverse problem of TDDFT on the torus]{Sobolev theory for the inverse problem of time-dependent density functional theory \\ on the torus}

\date{\today.}

\author[A.~B.~Lauritsen]{Asbjørn~Bækgaard~Lauritsen$^*$}
\email{$^*$lauritsen@ceremade.dauphine.fr}

\author[M.~Lewin]{Mathieu~Lewin$^\dagger$}
\email{$^\dagger$mathieu.lewin@math.cnrs.fr}

\author[J.~Oldenburg]{Jakob~Oldenburg$^\ddagger$}
\email{$^\ddagger$jakob.oldenburg@cnrs.fr}

\address{$^*$$^\dagger$$^\ddagger$CEREMADE, CNRS, Université Paris-Dauphine, PSL Research University, Place de Lattre de Tassigny, 75016 Paris, France}

\allowdisplaybreaks

\begin{document}

\begin{abstract}
We show that, given a sufficiently regular time-dependent density, there exists a unique potential such that the Schrödinger equation with this external potential has the prescribed one-particle density. We treat densities that are twice differentiable in time and of finite Sobolev regularity in space; the resulting potentials are continuous in time and of finite Sobolev regularity in space. 
Our method of proof is to rewrite the problem as a nonlinear Schrödinger equation with a loss of derivatives, which we in turn handle with a completion-of-the-square argument. 
This is a refinement of our recent work [Lauritsen, Lewin, and Oldenburg, arXiv:2609.08984], which handled this loss of derivatives using the Cauchy--Kovalevskaya theorem and treated analytic-in-space densities and potentials. 
\end{abstract}
\maketitle

\section{Introduction}
Time-dependent density functional theory (TDDFT) is one of the only viable methods of studying the dynamics of interacting systems such as molecules or solids \cite{Ullrichs-11,MarMaiNogGroRub-12,MarUllNogRubBurGro-06}. Its rigorous mathematical foundations are, in large parts, still open \cite{MAQUI_TDDFT-26,FouLamLewSor-16,Lampart-21,Penz-PhD,DFT-22}.

In TDDFT, a key role is played by the following inverse problem: Given a trajectory of  densities $\rho(t)$ (a non-negative function satisfying $\int\rho(t,x)\,\dx=N\in\N$), find an external time-dependent potential $V(t)$ such that the solution $\Psi(t)$ to the $N$-body Schrödinger equation with this potential has its one-particle density $\rho_{\Psi(t)} = \rho(t)$. This inverse problem, both with and without interaction between the electrons, lies at the heart of TDDFT and Kohn--Sham theory \cite{RunGro-84,vanLeeuwen-99, LauLewOld-V-26_ppt,MaiTodWooBur-10}. It has been studied numerically in \cite{RugGiePenLee-12,NieRugLee-13,RugPenLee-15,NieRugLee-18}.
The famous Runge--Gross theorem \cite{RunGro-84,FouLamLewSor-16,RugPenBau-JPA-09} states that, if it exists, this potential $V(t)$ is unique under the assumptions that it is real-analytic in time and smooth in space, whereas the wavefunction, density, and interaction are all smooth in space and time as well. However, it is expected that time-analyticity does not hold for physically interesting systems \cite{FouLamLewSor-16,YanMaiBur-12,YanBur-13,MaiTodWooBur-10}.
In \cite{LauLewOld-V-26_ppt,LauLewOld-V-short-26_ppt}, we gave the first proof of the existence and uniqueness of this potential for densities that are $C^2$ in time and analytic in space.

In the present paper, we refine the methods of \cite{LauLewOld-V-26_ppt} to cover densities that are of finite Sobolev regularity in space instead of analytic, see \cref{thm.main} below for the precise statement. 
More concretely, the inverse problem can be rewritten as a nonlinear Schrödinger equation with a loss of derivatives \cite{MaiTodWooBur-10,LauLewOld-V-26_ppt,LauLewOld-V-short-26_ppt}. In \cite{LauLewOld-V-26_ppt}, this loss of derivatives was handled using the Cauchy--Kovalevskaya theorem \cite{Nishida-77,Nirenberg-72,BaoGou-77,BaoGou-77b}. In the present paper, we rewrite the problem as a slightly different nonlinear Schrödinger equation, which still has a loss of derivatives and we handle it with a completion-of-the-square argument.

\medskip

\paragraph{\textbf{The paper is organized as follows.}} 
First, in \cref{sec.main.thm}, we state our main result and explain the key ideas of the proof and the novelties compared to \cite{LauLewOld-V-26_ppt}. Then, in \cref{sec.proof}, we give the proof of our results.

\medskip

\paragraph{\textbf{Acknowledgment.}} This work has benefited from French State support managed by ANR under the France 2030 program through the MaQuI CNRS Risky and High-Impact Research program (RI)$^2$ (grant agreement ANR-24-RRII-0001). We thank \'Eric Cancès, Théo Duez, Jari van Gog, and Julien Toulouse for stimulating discussions within this project.

\medskip
\paragraph{\textbf{AI declaration.}}
The crucial ideas of proof for the main theorem of the present paper were suggested by Large Language Models (LLMs). They have been completely rewritten by the authors and the paper contains no LLM output. More concretely, we gave as input to the LLM our paper \cite{LauLewOld-V-26_ppt} and asked to explore extensions of our work.
We used both Claude Opus~5.5 and GPT-6.1~Sol in alternation. 
Several proofs were found but the first ones were rather technical and hard to follow. (The first used analytic approximation and explicitly the solutions constructed in \cite{LauLewOld-V-26_ppt}.) 
After several iterations, involving in particular the suggestion to modify the nonlinear potential, a simple key idea was found. This idea, described in \cref{rmk.overview.ideas} below, is apparently well known in fluid mechanics. 
The proof based on this idea is the one we present in this paper.
Finally, the same LLMs were used at the final stage of writing the manuscript for proofreading.

\section{Main result}\label{sec.main.thm}
We consider the $D$-dimensional box $\Lambda = 	[0,1]^D$ with periodic boundary conditions and always assume that $D \leq 3$. For $N\geq1$, we consider the Schrödinger equation
\begin{equation}
\left\{
\begin{aligned}
i\partial_t\Psi(t) & = \cH_{V(t)}\Psi(t), \\
\Psi(0) & =\Psi_0,
\end{aligned}
    \label{eq:Schrodinger}
\right.
\end{equation}
with initial state $\Psi_0 \in L^2_a(\Lambda^N) = \bigwedge^N L^2(\Lambda)$ and the Hamiltonian 
\begin{equation*}
    \cH_{V} := \sum_{j=1}^N -\Delta_{x_j} + \sum_{j=1}^N V(x_j) + \sum_{1 \leq j < k \leq N} V_{ee}(x_j-x_k).
\end{equation*}
In particular, $\cH_0$ denotes the Hamiltonian without external potential. 
The interaction $V_{ee}$ is assumed to satisfy
\begin{assumption}[Interaction]\label{assumption:Vee}
We assume that the interaction potential $V_{ee}$ is real-valued, even, and satisfies the quadratic form bound
\begin{equation*}
V_{ee}^2 \leq C (1-\Delta).
\end{equation*}
Without loss of generality (otherwise shift $V_{ee}$ by a constant) we assume that $\cH_0\geq 1$. 
\end{assumption}

Taking the expectation against the constant function, it follows that $V_{ee}\in L^2(\Lambda)$ and, by the Rellich--Kato theorem \cite[Thm.~6.2]{Lewin-Spectral}, that $\cH_0$ is a self-adjoint operator with domain $H^2$ and form-domain $H^1$. Here we used the notation 

\begin{notation}[Sobolev spaces]
We denote by $H^s$ the standard Sobolev spaces, and use the following convention for their norms: 
\begin{equation*}
    \norm{f}_{H^s}^2 
    = \norm{f}_2^2 + \norm{(-\Delta)^{s/2}f}_2^2
    = \sum_{k\in 2\pi \Z^D} (1+|k|^{2s}) |\hat f(k)|^2,
    \qquad s \geq 0, 
\end{equation*}
where $\norm{\cdot}_p$ denotes the $L^p$-norm and we use the same notation $H^s$ for both the one- and many-body Sobolev spaces, whose norm is defined analogously. 
The Fourier coefficients of a function $f$ are normalized as $\hat f(k) = \int_\Lambda f(x) e^{-ikx} \, \rd x$.
\end{notation}

\begin{remark}[Coulomb]
We note that \cref{assumption:Vee} is satisfied for the 3D periodic Coulomb interaction given by the Fourier coefficients $\hat V_{ee}(k) = 4\pi /|k|^2$ for $k\in 2\pi \Z^3\setminus\{0\}$. 
\end{remark}

\begin{remark}[Direct problem]\label{rmk.direct}
In our case, the external potential $V$ will be of regularity $C^0([0,T];H^2(\Lambda))$ in which case it is well-known that,  for any initial state $\Psi_0 \in H^2(\Lambda^N)$, \eqref{eq:Schrodinger} admits a unique solution $\Psi\in C^0([0,T];H^2(\Lambda^N)) \cap C^1([0,T];L^2(\Lambda^N))$, see \cite{Yajima-87,ReeSim2} (or take $\sigma=0$ in \cite[Lem.~4.4]{LauLewOld-V-26_ppt} for an explicit proof).
\end{remark}

\begin{definition}[Spaces of wavefunctions and densities]
We define the spaces of admissible wavefunctions and densities as follows: 

\begin{itemize}[leftmargin=*]
\item 
We define the \emph{total momentum operator} $\cP$ by 
$$\cP=\sum_{j=1}^N -i \nabla_{x_j}.$$

\item
We define the following energies with a total momentum weight
\begin{equation}
	\label{eq:def_energy}
    E_0(\Psi) = \la \Psi, \cH_0  \Psi\ra, \quad 
	E_s (\Psi) = \la \Psi, \cH_0  \Psi\ra
	+\la \abs{\cP}^s \Psi, \cH_0 \abs{\cP}^s \Psi\ra \textrm{ for } s>0.
\end{equation}
We denote the corresponding norm $\norm{\Psi}_{H^1_s}= E_s (\Psi)^{1/2}$ and spaces
\begin{equation*}
    \boxed{H^1_s = \{ \Psi \in L^2(\Lambda^N) \ : \ \norm{\Psi}_{H^1_s} < \infty \}=D\big(\cH_0^{1/2}(1+|\cP|)^s\big).}
\end{equation*}
Note that the $H_s^1$-norm is equivalent to $\norm{(1+|\cP|^s)\cdot}_{H^1}$ since $\cP$ and $\cH_0$ commute and by \cref{assumption:Vee}. 

\smallskip

\item
Our space of permissible density-trajectories is
\begin{equation*}
    \boxed{\cD = C^0([0,T];H^{4})\cap C^1([0,T];H^{2})\cap C^2([0,T];L^2).}
\end{equation*}
(we do not emphasize the $T$-dependence in our notation) and it is endowed with the usual norm 
\begin{equation*}
	\norm{\rho}_{\cD}
 =\max_{0\leq t \leq T}
 \big(
 \norm{\rho}_{H^{4}}
 +\norm{\partial_t\rho}_{H^{2}}
 +\norm{\partial_t^2\rho}_{2}\big).
\end{equation*}
\end{itemize}
\end{definition}

To state the main theorem, we recall that the \emph{density} $\rho_\Psi$ and \emph{current} $j_\Psi$ of a wavefunction $\Psi$ are given by 
\begin{align*}
\rho_\Psi(x)
& := 
N  \idotsint_{\Lambda^{N-1}} \abs{\Psi(x,x_2,...,x_N)}^2 \,\dx_2\cdots\dx_N
\\
j_{\Psi}(x)
& := 
2N\Im \idotsint_{\Lambda^{N-1}}\overline{\Psi(x,x_2,...,x_N)} \nabla_x\Psi(x,x_2,...,x_N)\,\dx_2\cdots\dx_N.
\end{align*}
Moreover, we recall that any solution $\Psi(t)$ to the Schrödinger equation \eqref{eq:Schrodinger} satisfies the \emph{continuity equation}
\begin{equation}
    \partial_t \rho_{\Psi(t)} + \nabla \cdot j_{\Psi(t)} = 0. 
    \label{eq:continuity_equation}
\end{equation}
Our main theorem then reads

\begin{theorem}[The inverse problem]\label{thm.main}
Let $V_{ee}$ satisfy \cref{assumption:Vee}. 
Let $T>0$ and $\Psi_0\in H^2 \cap H^1_2$ be the initial (anti-)symmetric wavefunction with $\norm{\Psi_0}_2=1$.
Consider a time-dependent density $\rho \in \cD$ for which we have
\begin{itemize}
    \item normalization:  $\int_\Lambda \rho(t,x) \, \rd x = N$ for all $0\leq t\leq T$,
    \item initial compatibility:  $\rho(0)=\rho_{\Psi_0}$ and $\partial_t\rho(0)=-\nabla\cdot j_{\Psi_0}$, and 
	\item positivity:  $\rho(t) \geq c > 0$ for all $0\leq t\leq T$. 
\end{itemize}
Then there exists $0 < T' \leq T$ and a unique real-valued potential $V\in C^0([0,T'];H^2)$ with $\int V(t,x) \, \rd x = 0$ such that the solution $\Psi$ of the time-dependent Schrödinger equation \eqref{eq:Schrodinger} reproduces the density $\rho_{\Psi(t)} = \rho(t)$ for all $0\leq t \leq T'$. 
\end{theorem}

\begin{remark}[Coulombic systems are not covered]
We remark that the Coulomb potential $V$ generated by point nuclei is  unbounded and thus not in $H^2$. Moreover, 
ground state densities of Coulombic systems in 3D with point nuclei have cusps at the nuclei \cite{FouHofOstHofOstSor-07} and are not in $H^4$. Concretely, the ground state density of a (periodic) Hydrogen atom behaves as $e^{-|x|} \notin H^4_{\rm loc}$ near the nucleus. 
\end{remark}

\begin{remark}[Initial regularity]
Note that the assumption of initial compatibility with a density $\rho\in \cD$ requires more regularity properties (namely $\rho_{\Psi_0}\in H^4$ and $\nabla \cdot j_{\Psi_0} \in H^2$) than what can be obtained from the sole assumption that $\Psi_0\in H^2\cap H_2^1$. 
\end{remark}

\begin{remark}[Spin]
For simplicity we consider here spinless fermions or bosons. The theorem extends easily to include spin with $\rho$ being the spin-summed density.  
\end{remark}

\begin{remark}[Relevance for TDDFT]
\cref{thm.main} is important for TDDFT, but not the final answer. In TDDFT one needs a one-to-one map between two spaces of potentials and densities~\cite{GroDobPet-96,Ullrichs-11,MAQUI_continuous_long-26,FouLamLewSor-16}, and this is not furnished by Theorem~\ref{thm.main}. For an external potential $V\in C^0([0,T];H^2)$, the associated time-dependent density $\rho_\Psi$ of the solution $\Psi$ to the Schrödinger equation \eqref{eq:Schrodinger} does not necessarily belong to $\cD$. 
Obviously we have a one-to-one map between $\cD$ and the set of inverse potentials in $C^0([0,T'];H^2)$ constructed in \cref{thm.main}. Giving a simple direct characterization of this set is a delicate problem, see \cite{Lampart-21}. 
\end{remark}

\begin{remark}[{Overview of ideas and comparison to \cite{LauLewOld-V-26_ppt}}]
\label{rmk.overview.ideas}
In \cite{LauLewOld-V-26_ppt} the main ideas were: 
\begin{enumerate}[1.]
    \item Use the total momentum $\cP$ as the derivative operator/weight measuring regularity of the wavefunction (inspired by \cite{Hunziker-86}) instead of regularity in all direction as would come from the Sobolev space $H^s(\Lambda^N)$. Require only $H^2(\Lambda^N)$ (the domain of $\cH_0$) for the other directions. 
    \item Following the suggestion of~\cite{MaiTodWooBur-10}, recast the problem as a nonlinear Schrödinger equation in $\Psi$, with a nonlinear potential $V\rightsquigarrow V_\Psi$ (defined in \eqref{eq:VPsi} below) carefully constructed to lose only one derivative.
    \item Estimate the derivative loss in terms of the total momentum only, or only using $H^2$ in the other directions. 
    \item Apply the Cauchy--Kovalevskaya technique to find solutions.
\end{enumerate}
The main novelty of the present paper is the refinement of the second idea by introducing the new nonlinear potential $V\rightsquigarrow U_\Psi$ 
(defined in \eqref{eq:UPsi} below), which allows one to \textbf{fully absorb all losses of derivatives by a completion-of-the-square argument}. More precisely, we use the desired density constraint to insert in the nonlinear equation an additional term which vanishes exactly for the sought-after solution, but can serve as a regulator to control the lost  derivatives in the nonlinear equation. This allows us to replace the Cauchy--Kovalevskaya machinery of Step 4 by a simpler Grönwall energy estimate. The regulator is just based on the continuity equation~\eqref{eq:continuity_equation}, as we explain in detail below.
\end{remark}

\subsection{Ideas of the proof}
The main part of \cref{thm.main} is the existence of such a potential $V$; the uniqueness is a comparatively simpler separate argument, and we do not discuss it in this subsection. Thus, one needs to construct a candidate potential $V$ and show that the associated solution $\Psi(t)$ has the desired density, $\rho_{\Psi(t)}=\rho(t)$.

To find this candidate potential, the idea is to consider an equation it \emph{would have to satisfy} to reproduce the density $\rho$. 
By studying the direct problem as in \cite{LauLewOld-V-26_ppt,LauLewOld-V-short-26_ppt}, one finds that the density and potential are linked via the \emph{force-balance equation} \cite{vanLeeuwen-99}, which leads one to define the nonlinear potential $V_\Psi$ as 
\begin{equation}\label{eq:VPsi}
    V_\Psi(t) : = \frac{1}{2} K_{\rho(t)}^{-1}\left(-\partial_t^2\rho(t) - \Delta^2\rho(t) + Q_{\Psi}\right) ,
\end{equation}
where $K_{\rho} f = - \nabla\cdot(\rho\nabla f)$ and $Q_\Psi = Q_{\Psi,\Psi}$ with (denoting $\partial^\alpha = \partial_{x^\alpha}$ for the derivative in the $\alpha$'th coordinate of $x=(x^1,\ldots,x^D)$, $X=(x_2,\ldots,x_N)$, and $Y = (x_3,\ldots, x_N)$)
\begin{align}
    Q_{\Psi_1,\Psi_2}(x) 
    & = 4N\sum_{\alpha,\beta=1}^D \partial^\alpha\partial^\beta\int \rd X \,\overline{\partial^\alpha \Psi_1(x,X)} \partial^\beta\Psi_2(x,X)
    \nn \\* & \quad 
	+ 2N(N-1)\Delta_x  \iint \rd y \, \rd Y \, \overline{\Psi_1(x,y,Y)}\Psi_2(x,y,Y) V_{ee}(x-y)
    \nn \\* & \quad 
    -2 N(N-1) \sum_{\alpha=1}^D \partial^\alpha \iint \rd y \, \rd Y \,  \partial^\alpha \left(\overline{\Psi_1(x,y,Y)}  \Psi_2(x,y,Y)\right)    V_{ee}(x-y).
    \label{eq:Q_Psi1_Psi2}
\end{align} 
In~\eqref{eq:VPsi} we have replaced all the $\rho_\Psi$ that would normally appear in the force balance equation by the given $\rho$ so that $\Psi$ only appears in $Q_\Psi$. This is particularly important for the  term $\Delta^2\rho$ that would otherwise involve four derivatives of $\Psi$. Since we have assumed that $\rho$ is strictly positive, the operator $K_\rho$ is invertible on mean-zero functions (\cref{lem.Krho} below), and thus $V_\Psi$ is well-defined.

One is then led to study the nonlinear $N$-particle Schrödinger equation  \cite{MaiTodWooBur-10,LauLewOld-V-26_ppt,LauLewOld-V-short-26_ppt}
\begin{equation}\label{eq:SEnonlinearV}
    i\partial_t \Psi (t) 
    = \cH_{V_{\Psi(t)}(t)}\Psi(t), 
    \qquad \Psi(0) = \Psi_0 . 
\end{equation}
In \cite{LauLewOld-V-26_ppt} we showed that \eqref{eq:SEnonlinearV} admits a (unique) solution $\Psi_{\rm NLS}$ in the class of functions that are analytic in the center-of-mass direction. From this solution, the sought-after potential in \cref{thm.main} is then $V = V_{\Psi_{\rm NLS}}$ and it remains to check that the density is reproduced. The nonlinear potential $V_\Psi$ morally has a loss of one derivative; the function $Q_\Psi$ involves three spatial derivatives of $\Psi$, and the inversion of the elliptic operator $K_\rho$ gains two. 
This loss of one derivative permits one to use the Cauchy--Kovalevskaya theorem \cite{Nirenberg-72,Nishida-77,BaoGou-77,BaoGou-77b} to conclude the existence of solutions in the analytic-in-space class, which is what we did in \cite{LauLewOld-V-26_ppt}.

The key new idea of this paper, that was suggested to us by LLMs, is to instead solve the $N$-particle Schrödinger equation with the nonlinear potential 
\begin{equation}\label{eq:UPsi}
	\boxed{U_{\Psi}(t) : = V_\Psi (t) -(\nabla\cdot j_{\Psi}+\partial_t \rho).}
\end{equation}
This nonlinear potential now has a loss of two spatial derivatives. However, crucially, the loss has a definite sign and acts as a regularization. (More precisely the sign emerges in the time-derivative of the energy $E_2$ in~\eqref{eq:def_energy}, which can then be proved to  stay bounded by a Grönwall argument, see also \cref{rmk.illustration.viscosity.N=1,rmk.illustration.N=1}.) 
The added term $-\nabla\cdot j_\Psi - \partial_t\rho$ vanishes at consistency $\rho=\rho_\Psi$ by the continuity equation \eqref{eq:continuity_equation}. 
Similar ideas of adding a regularizing term to the equation which vanishes for solutions have been used for the numerical analysis of fluid equations \cite{OlsReu-04,LiuLiuPeg-07},
see also \cite{FouresBruhat-52,GunCalHinMar-05,HinVas-18,HafHinVas-20} for related ideas in the context of general relativity. 
In addition, a similar modification with $-\nabla\cdot j_\Psi - \partial_t\rho$ was used in an iterative scheme for the numerical study of the present inverse problem of TDDFT in \cite[Sec.~6.2]{NieRugLee-18}. 
Finally, we could instead consider an added term $-\nu (\nabla\cdot j_\Psi + \partial_t\rho)$ for any $\nu > 0$ and the same arguments would apply. For simplicity we take $\nu=1$.

To solve the nonlinear Schrödinger equation 
\begin{equation}
    i\partial_t \Psi (t) 
    = \cH_{U_{\Psi(t)}(t)}\Psi(t), 
    \qquad \Psi(0) = \Psi_0 
    \label{eqn.nonlinear.SE.U}
\end{equation}
we employ classical techniques based on \emph{a priori} estimates involving  the Lyapounov functional $E_2$ in~\eqref{eq:def_energy}. Namely we first regularize the equation using a spectral cut-off $L$ for $\cH_0$, which reduces the problem to finite-dimension, where existence of solutions follows by standard tools of ODEs. For this spectrally localized problem we show that the energy $E_2(\Psi_L)$ is uniformly bounded in the spectral localization $L$, using a Grönwall estimate and a completion-of-the-square argument to handle the loss of derivatives. See \cref{rmk.illustration.viscosity.N=1,rmk.illustration.N=1} below for an illustration in the one-particle setting. Passing to the limit $L\to\ii$ we use the boundedness of $E_2$ to find a solution $\Psi_{\rm NLS}$ to the nonlinear equation \eqref{eqn.nonlinear.SE.U}, and thus a candidate potential $V=U_{\Psi_{\rm NLS}}$. Finally, we check that this potential $V$ reproduces the desired density $\rho$. Uniqueness is obtained using instead the energy $E_0$ in~\eqref{eq:def_energy}.

\begin{remark}[The additional term has a regularization effect]
\label{rmk.illustration.viscosity.N=1}
To explain why the additional term $-(\partial_t\rho+\nabla\cdot j_\Psi)$ in the potential acts as a regulator, let us look at the simpler equation for $N=1$
\begin{equation}
i\partial_t \psi=(-\Delta+V_{\rm ext}(t)-\nabla\cdot j_\psi)\psi, 
\label{eq:simple_viscosity}
\end{equation}
where $j_\psi=2\Im(\overline{\psi}\nabla \psi)$ and $V_{\rm ext}(t)$ is a given nice external potential (we think that it contains the term $-\partial_t\rho$ involving the prescribed density). Here we denoted the wavefunction by $\psi$ to distinguish it from the many-body case. In the hydrodynamics formulation $\psi=\sqrt{\rho_\psi} e^{i\theta}$, we have $j_\psi=2\rho_\psi\nabla\theta$ and the equation becomes
$$\begin{cases}
    \partial_t\rho_\psi+2\nabla\cdot(\rho_\psi\nabla\theta)=0\\
    \partial_t\theta+|\nabla\theta|^2-2\nabla\cdot(\rho_\psi\nabla\theta)+V_{\rm ext}-\frac{\Delta\sqrt{\rho_\psi}}{\sqrt{\rho_\psi}}=0.
\end{cases}$$
We see that the additional term $-2\nabla\cdot(\rho_\psi\nabla\theta)$ in the second equation acts as a viscosity on the phase $\theta$.
\end{remark}

\begin{remark}[Heuristics for the completion-of-the-square argument]
\label{rmk.illustration.N=1}
To illustrate the completion-of-the-square argument, we then add to the external potential $V_{\rm ext}$ in \eqref{eq:simple_viscosity} a nonlinear potential $|\nabla \psi|^2$, which captures the loss of one derivative. 
The nonlinear equation now reads
\begin{equation*}
    i\partial_t \psi = (-\Delta + U_\psi)\psi,
    \quad
    U_\psi = V_{\rm ext}(t)+|\nabla \psi|^2 -\nabla\cdot j_{\psi}.
\end{equation*}
Consider the growth of the kinetic energy 
\begin{align*}
    \partial_t \norm{\psi}_{H^1}^2 
    &= \partial_t \norm{\nabla \psi}_{2}^2
    = 2\Re \pscal{\nabla\psi, \nabla (i\Delta - i U_\psi) \psi}
     = 2 \Im \pscal{\nabla \psi, \nabla U_\psi \psi }\\
      &=2 \Im \int \overline{\nabla \psi}\psi \nabla U_\psi  
      =\int\nabla \cdot j_\psi \, U_\psi 
      =\int\nabla \cdot j_\psi  \left(|\nabla \psi|^2 -\nabla\cdot j_{\psi}+V_{\rm ext}\right)\\
      &\leq - \norm{\nabla \cdot j_\psi}_2^2 
      + \norm{\nabla \cdot j_\psi}_2\left( \norm{V_{\rm ext}}_2
      + \norm{\nabla \psi}_4^2\right).
\end{align*}
This computation illustrates the mechanism of how the term with the most derivatives appears as a negative square and could absorb the loss of derivatives. Unfortunately, for the kinetic energy ($H^1$-norm) we face an issue with the term $\norm{\nabla\psi}_4^2$, which we cannot bound by the $H^1$-norm of $\psi$ and close the Grönwall estimate. 
This is precisely the reason we work with $H^1_2$ for $\Psi$, which corresponds to $H^3$ for $\psi$. For the $H^3$-norm we find instead the bound (using a similar bound for the linear terms)
\begin{equation*}
    \partial_t \norm{\psi}_{H^3}^2 \leq - \frac{1}{2} \norm{\nabla \cdot j_\psi}_{H^2}^2 + C\norm{V_{\rm ext}}_{H^2}^2 + C\norm{\psi}_{H^3}^4 
\end{equation*}
see \eqref{eqn.bdd.partial.t.E2} below. The Grönwall argument thus closes and we conclude the boundedness of solutions in $H^3$, at least until the blowup-time of the equation $\partial_t z = 1 + z^2$.
\end{remark}

\section{Proof}\label{sec.proof}
We now give the proof of \cref{thm.main}. The main part is to show that the nonlinear equation \eqref{eqn.nonlinear.SE.U} admits a (unique) solution in some appropriate space. 
Using this, we give finally the proof of \cref{thm.main} in \cref{sec.final.steps.main.thm} below.

\subsection{Preliminaries}
We give some preliminary bounds, which are analogues of bounds from \cite{LauLewOld-V-26_ppt}. 
First, we introduce some notation that will be used throughout the proof. 

\begin{notation}[Further notation]
We introduce the following quantities
\begin{itemize}[leftmargin=*]
    \item 
We use the Japanese bracket notation 
\begin{equation*}
    \pscal{p} = (1+|p|^2)^{1/2} 
\end{equation*}
for any vector $p\in \R^n$ and any $n$. 

\item 
We define the $N$-body operators (for any potential $V$)
\begin{equation*}
    \cV = \sum_{j=1}^N V(x_j), \qquad 
    \cT = \sum_{j=1}^N -\Delta_{x_j}, \qquad 
    \cV_{ee} = \sum_{1 \leq j < k \leq N} V_{ee}(x_j - x_k).
\end{equation*}
In particular we write $\cV_\Psi = \sum_{j=1}^N V_\Psi(x_j)$ and $\cU_\Psi = \sum_{j=1}^N U_\Psi(x_j)$. 

\item 
Define the bilinear density as 
\begin{equation*}
\rho_{\Psi_1,\Psi_2}(x) 
=N \idotsint \overline{\Psi_1(x,x_2,...,x_N)} \Psi_2(x,x_2,...,x_N)\,\dx_2\cdots\dx_N.     
\end{equation*} 
\end{itemize}
\end{notation}

\begin{lemma}[{Invertibility of $K_\rho$}]
\label{lem.Krho}
Let $\rho \in \cD$ with $\rho\geq c > 0$. 
Then, $K_{\rho(t)}$ is invertible on the space $H^{s,\perp} = \{f\in H^s : \int_\Lambda f = 0\}$ for 
$s \in \{0,2\}$. 
Moreover,
we have uniformly in time (with $\norm{\cdot}_{\mathrm{op},s}$ denoting the operator norm on $H^{s,\perp}$)
\begin{equation*}
\norm{K_{\rho(t)}^{-1}(1-\Delta)}
_{\mathrm{op},s} \leq C, 
\qquad 
\norm{\left(K_{\rho(t)}^{-1} -K_{\rho(t')}^{-1}\right)(1-\Delta)}_{\mathrm{op},s}
    \leq  C 
    \norm{\rho(t)-\rho(t')}_{H^3}.
\end{equation*}
\end{lemma}

By a simple interpolation argument, we get the same bounds for all $0\leq s \leq 2$. As we shall not need it, we omit the details. 

\begin{proof}
The case $s=2$ is proved as in \cite[Lem.~6.7]{LauLewOld-V-26_ppt} (choose the analyticity parameter $\sigma=0$ there).
The case $s=0$ follows from the $s=2$ case as follows. By self-adjointness, $\snorm{K_{\rho(t)}^{-1}(1-\Delta)}_{\mathrm{op},0} = \snorm{(1-\Delta)K_{\rho(t)}^{-1}}_{\mathrm{op},0}$. Then, with $g = (1-\Delta)^{-1} f$, 
\begin{equation*}
    \norm{(1-\Delta)K_{\rho(t)}^{-1}f}_2 
    \leq C\norm{K_{\rho(t)}^{-1}(1-\Delta) g}_{H^2}
    \leq C \norm{K_{\rho(t)}^{-1}(1-\Delta)}_{\mathrm{op},2} \norm{f}_2. 
\end{equation*}
The same argument applies for the difference $K_{\rho(t)}^{-1} - K_{\rho(t')}^{-1}$. 
\end{proof}

\begin{lemma}[Asymmetric bilinear density bound]
\label{lem:bilinear}
For any $\Psi_1,\Psi_2\in L^2$ and any $0 \leq s \leq s_1, s_2$ with $s_1 + s_2 > s + D/2$ we have
\begin{align*}
\norm{\rho_{\Psi_1,\Psi_2}}_{H^s} 
& \leq C 
\norm{\pscal{\cP}^{s_1}\Psi_1}_2 \norm{\pscal{\cP}^{s_2}\Psi_2}_2.
\end{align*}
In particular, 
$\norm{\rho_{\Psi_1,\Psi_2}}_{H^s} \leq C \norm{\pscal{\cP}^2\Psi_1}_2 \norm{\pscal{\cP}^s\Psi_2}_2$ for $s\in \{0,1,2\}$.
\end{lemma}

\begin{proof}
This follows by an adaptation of \cite[Lem.~4.5]{LauLewOld-V-26_ppt}. We sketch the details. 
Denote $\vec p = (p_1,\ldots, p_N)$, $p+\vec p = (p+p_1,p_2,\ldots, p_N)$, and $P = \sum_{j=1}^N p_j$. Then, Fourier coefficients are given by $\hat \rho_{\Psi_1,\Psi_2}(p) = N \sum_{p_j \in 2\pi \Z^D} \overline{\hat \Psi_1}(\vec p) \hat\Psi_2(p + \vec p)$. 
First, we bound 
$\pscal{p}^s \leq C \pscal{P}^s + C \pscal{p+P}^s$. 
To bound this further, we note that for any $z_1,z_2 >0$ and $a_1,a_2 \geq0$ we have $z_j^{-a_j} \leq \min\{z_1,z_2\}^{-a_j}$. Thus, 
\begin{equation}\label{eqn.product.powers}
z_1^{-a_1}z_2^{-a_2} \leq \min\{z_1,z_2\}^{-a_1-a_2} \leq z_1^{-a_1-a_2} + z_2^{-a_1-a_2}.
\end{equation}
Using this we have 
\begin{multline*}
    \pscal{P}^s = \pscal{P}^{s_1} \pscal{p+P}^{s_2} \pscal{P}^{-(s_1-s)} \pscal{p+P}^{-s_2}
    \\*
    \leq \pscal{P}^{s_1} \pscal{p+P}^{s_2} \left(\pscal{P}^{-(s_1+s_2-s)} + \pscal{p+P}^{-(s_1+s_2-s)}\right).
\end{multline*}
Similarly, the same bound holds for $\pscal{p+P}^s$. (This uses $s_2\geq s$.) We conclude
\begin{multline*}
    \pscal{p}^{s} |\hat \rho_{\Psi_1,\Psi_2}(p)| 
    \leq C \sum_{p_j\in 2\pi \Z^D} \left(\pscal{P}^{-(s_1+s_2-s)} + \pscal{p+P}^{-(s_1+s_2-s)}\right) 
    \\* 
    \times \pscal{P}^{s_1} |\hat \Psi_1(\vec p)|  \pscal{p+P}^{s_2}  |\hat \Psi_2(p + \vec p)|. 
\end{multline*}
Computing the $p$-sum and using the Cauchy--Schwarz inequality as in \cite[Lem.~4.5]{LauLewOld-V-26_ppt} we conclude the desired, noting that $s_1+s_2 > s + D/2$ ensures that $\sum_p \pscal{p}^{-2(s_1+s_2-s)} < \infty$. 
\end{proof}

\begin{lemma}[One-body products]\label{lemma:embeddings}
For $\Psi\in L^2(\Lambda^N), f\in L^2(\Lambda)$ and any $0 \leq s \leq s_1$ and $s_2\geq 0$ with $s_1 + s_2 > D/2$ we have (by a slight abuse of notation $(f\Psi)(x,X) = f(x) \Psi(x,X)$)
\begin{equation*}
    \norm{f\Psi}_{H^s} \leq C \norm{f}_{H^{s_1}} \norm{\pscal{\cP}^{s_2}\Psi}_{H^s}.
\end{equation*}
In particular,
$\norm{f\Psi}_{2} \leq C \norm{f}_{H^1} \norm{\pscal{\cP}\Psi}_2$ and
$\norm{f\Psi}_{H^s} \leq C \norm{f}_{H^s} \norm{\pscal{\cP}^{2-s}\Psi}_{H^s}$ for $s\in \{0,1,2\}$.
\end{lemma}

\begin{proof}
This follows by an adaptation of \cite[Lem.~4.2]{LauLewOld-V-26_ppt} with similar modifications as in the proof of \cref{lem:bilinear}. We give a sketch. Denoting $\vec p = (p_1,\ldots,p_N)$, $\vec p - p = (p_1-p,p_2,\ldots,p_N)$, and $P=\sum_{j=1}^N p_j$ we have $\hat {f\Psi}(\vec p) = \sum_{p\in 2\pi \Z^D} \hat f(p) \hat \Psi(\vec p - p)$. Next, $ \pscal{\vec p}^s \leq C \pscal{\vec p - p}^s + C \pscal{p}^s$. Using that $\pscal{P} \leq C \pscal{\vec p}$ and \eqref{eqn.product.powers}, we have
\begin{equation*}
    \pscal{\vec p - p}^s + \pscal{p}^s \leq C \pscal{p}^{s_1} \pscal{P-p}^{s_2} \pscal{\vec p - p}^s \left(\pscal{p}^{-s_1-s_2} + \pscal{P - p}^{-s_1-s_2}\right). 
\end{equation*}
With the discrete Young and Cauchy--Schwarz inequalities as in \cite[Lem.~4.2]{LauLewOld-V-26_ppt} we conclude the desired, noting that $s_1+s_2>D/2$ ensures $\sum_p \pscal{p}^{-2s_1-2s_2} < \infty$.
\end{proof}

\begin{lemma}[The potential $V_\Psi$]
\label{lemma:VPsi}
Let $V_\Psi$ be as defined in \eqref{eq:VPsi}. Then, 
\begin{align}
  \norm{V_{\Psi}}_{H^2} 
  & \leq C \norm{\Psi}_{H^1_2}^2 + C \norm{\rho}_{\cD}
  , 
  \label{eqn.bdd.VPsi.Hs}
    \\
    \norm{V_{\Psi_1}-V_{\Psi_2}}_{H^s} 
  & \leq C \norm{\Psi_1-\Psi_2}_{H^1_{s}}\norm{\Psi_1+\Psi_2}_{H^1_{2}}, 
  \qquad s \in \{0,2\}
    \label{eqn.bdd.VPsi1-VPsi2.Hs}
    \\
    \norm{V_{\Psi}(t)-V_{\Psi}(t')}_{H^2}
     & \leq C 
     \left( \norm{\rho(t) - \rho(t')}_{H^4} + \norm{\partial_t^2 \rho(t) - \partial_t^2\rho(t')}_2\right)
     \left(
     1
     + \norm{\Psi}_{H^1_2}^2
     \right)
     .
    \label{eqn.VPsi.cts.in.time}
\end{align}
\end{lemma}

\begin{proof}
This follows by an adaptation of \cite[Lems.~5.2 and 6.13]{LauLewOld-V-26_ppt}. We give the details. 
We rewrite \eqref{eq:Q_Psi1_Psi2} in a form containing the interaction operator $\cV_{ee}$ as
\begin{equation*}
    Q_{\Psi_1, \Psi_2} = 4 \sum_{\alpha,\beta=1}^D \partial^\alpha \partial^\beta \rho_{\partial^\alpha \Psi_1, \partial^\beta \Psi_2}
    + 2 \Delta \rho_{\Psi_1, \cV_{ee}\Psi_2} - 2 \sum_{\alpha=1}^D \partial^\alpha \left( \rho_{\partial^\alpha \Psi_1, \cV_{ee}\Psi_2} + \rho_{ \cV_{ee}\Psi_1, \partial^\alpha\Psi_2}\right). 
\end{equation*}
Here we used the fermionic symmetry as well as a cancellation from contributions of $V_{ee}(x_j-x_k)$ for $j,k>1$.
Next, note that \cref{assumption:Vee} applied in one variable, with all others fixed gives the bound $\norm{\cV_{ee}\Psi}_2 \leq C \norm{\Psi}_{H^1}$. 
Now, with \cref{lem:bilinear} (and using that $\cP$ and $\cV_{ee}$ commute) we thus conclude the bound 
\begin{equation*}
\norm{(1-\Delta)^{-1}Q_{\Psi_1,\Psi_2}}_{H^{s}} \leq C \norm{\pscal{\cP}^{s_1}\Psi_1}_{H^1} \norm{\pscal{\cP}^{s_2}\Psi_2}_{H^1} \leq C \norm{\Psi_1}_{H_{s_1}^1} \norm{\Psi_2}_{H_{s_2}^1}
\end{equation*}
for $s_1, s_2 \geq s$ with $s_1 + s_2 > s + D/2$.
Then, from \cref{lem.Krho} and the identity $V_{\Psi_1} - V_{\Psi_2} = \frac{1}{2} \Re K_{\rho}^{-1} Q_{\Psi_1 + \Psi_2, \Psi_1- \Psi_2}$ the claimed bounds in \eqref{eqn.bdd.VPsi.Hs} and \eqref{eqn.bdd.VPsi1-VPsi2.Hs} follow by choosing $s_2=s$ and $s_1=2$.

Lastly, to prove \eqref{eqn.VPsi.cts.in.time}, we compute
\begin{align*}
     & 2\norm{V_{\Psi}(t)-V_{\Psi}(t')}_{H^2}
     \\ & \quad =
     \norm{
     K_{\rho(t)}^{-1}\left(-\partial_t^2\rho(t) - \Delta^2\rho(t) + Q_{\Psi}\right) 
     -K_{\rho(t')}^{-1}\left(-\partial_t^2\rho(t') - \Delta^2\rho(t') + Q_{\Psi}\right)
     }_{H^2}
     \\ & \quad 
     \leq \norm{
     K_{\rho(t')}^{-1}(1-\Delta) 
     (1-\Delta)^{-1}\left(\partial_t^2\rho(t) + \Delta^2\rho(t) -\partial_t^2\rho(t') - \Delta^2\rho(t') \right)
     }_{H^2}
     \\ & \qquad 
     +\norm{
     \left( K_{\rho(t)}^{-1} 
     -K_{\rho(t')}^{-1} \right)(1-\Delta)  (1-\Delta)^{-1}\left(-\partial_t^2\rho(t) - \Delta^2\rho(t) + Q_{\Psi}\right)
     }_{H^2}
     \\ & \quad 
     \leq C 
     \left( \norm{\rho(t) - \rho(t')}_{H^4} + \norm{\partial_t^2 \rho(t) - \partial_t^2\rho(t')}_2\right)
     \left(
     1
     +\norm{\rho}_\cD 
     + \norm{\Psi}_{H^1_{2}}^2
     \right)
     , 
\end{align*}
where we used \cref{lem.Krho,lem:bilinear} as before.
\end{proof}

\subsection{Existence of regularized solutions}
We first show the existence of regularized solutions. 

\begin{lemma}[Regularized solutions]\label{lemma:regularized}
Let $V_{ee},\Psi_0, \rho, T$ be as in \cref{thm.main}.
Define the spectral projection
\begin{equation}\label{eq:def_spectral_projections}
    \Pi_L = \mathbbm{1}\left(\cH_0\leq L\right)
\end{equation}
and let $L > 0$ large enough such that $\norm{\Pi_L\Psi_0}_2 > 1/2$. 
Then the equation
\begin{equation}\label{eq:SEnonlinear_regularized}
i\partial_t \Psi_L (t) 
= \Pi_L\cH_{ U_{\Psi_L(t)}(t)}\Pi_L \Psi_L(t), 
\quad \Psi_L(0) = \frac{\Pi_L\Psi_0}{\norm{\Pi_L\Psi_0}_2} 
\end{equation}
admits a unique solution $\Psi_L\in C^1([0,T];\Ran \Pi_L)$, with $\Ran \Pi_L$ the range of $\Pi_L$. 
\end{lemma}

\begin{proof}
	Note that the projection $\Pi_L$ from \eqref{eq:def_spectral_projections} reduces the problem to a finite dimensional one since $\dim \Ran \Pi_L < \ii$ for any $L$ (since we work on a torus). 
	For $\Phi\in\Ran \Pi_L$ we have  $\norm{\pscal{\cP}^s\Phi}_{H^2}\leq C_{L,s} \norm{\Phi}_{2} $ for any $s\geq 0$. In particular, all relevant norms are equivalent to the $L^2$-norm.
	Local existence now follows by the Picard--Lindelöf theorem as follows. 
	We recall equation \eqref{eq:SEnonlinear_regularized}
		\begin{equation*}
    i\partial_t \Psi_L (t) 
    = \Pi_L \cH_{U_{\Psi_L(t)}(t)}\Pi_L \Psi_L(t)
    = \cH_{0}\Pi_L \Psi_L(t) +\Pi_L \cU_{\Psi_L(t)}(t)\Pi_L \Psi_L(t),
\end{equation*}
which, in particular, preserves $\Ran \Pi_L$.
The right hand side	of the equation is continuous in time (recall the definition of the potential in \eqref{eq:VPsi} and \eqref{eq:UPsi}) by \cref{lem.Krho} and since $\rho\in \cD$.
Moreover, by \cref{lemma:VPsi} we find that the right-hand-side is locally Lipschitz in $\Psi$ since it is a polynomial. 
Thus, by the Picard--Lindelöf theorem, we conclude the existence and uniqueness of solutions on some possibly shorter time-interval $[0,T']$. Either we have $T'=T$ or we have that some norm of $\Psi_L$ blows up at time $T'$. 
However, all norms are equivalent to the $L^2$-norm and by construction, $\norm{\Psi_L}_2=1$ is conserved. Thus, $T'=T$.
\end{proof}

\subsection{Energy estimates}
Next, we bound the energy $E_2$ of the regularized solutions. We prove

\begin{lemma}[Energy estimates]\label{lemma:energy}
Let $\Psi_L$ be the solution to \eqref{eq:SEnonlinear_regularized} from \cref{lemma:regularized}. There exists a time $0<T'\leq T$ such that, uniformly in $L$, we have 
\begin{align}    
\label{eq:bound_E_2}
	\sup_{0\leq t\leq T'} E_2(\Psi_L (t)) & \leq C, 
    \\
\label{eq:bound_j}
	\int_0^{T'} \rd t \, \norm{\nabla\cdot j_{\Psi_L(t)}}_{H^2}^2 & \leq C, 
    \\
\label{eq:bound_U}
	\int_0^{T'} \rd t \, \norm{U_{\Psi_L(t)}(t)}_{H^2}^2 
    & \leq C,
    \\
\label{eq:bound_H^2}
 \norm{\Psi_L(t)}_{H^2} & \leq C,
 \\
\label{eq:bound_partial_t_Psi}
 \norm{\partial_t\Psi_L(t)}_{2} & \leq C,
\end{align}
where the energy $E_2$ and potential $U_{\Psi_L}$ are as defined in \eqref{eq:def_energy} and \eqref{eq:UPsi}. 
\end{lemma}

The proof goes via a Grönwall estimate on $E_2(\Psi_L)$. For later purposes it will be useful to have a slightly more general bound on $\partial_t E_2(\Psi_L)$ (which exists as a classical derivative by \cref{lemma:regularized}). 
Using $\Pi_L\Psi_L=\Psi_L$, that $\Pi_L$, $\cH_0$, and $\cP$ all commute, and that $\cH_0$ is a self-adjoint operator, we find
\begin{align}
    \partial_t E_2(\Psi_L)
    & = 2 \Im \pscal{\Psi_L, \cH_0 (\cH_0 + \Pi_L \cU_{\Psi_L}\Pi_L)\Psi_L}
    \nn
    \\ & \quad 
    + 2 \Im \pscal{\cP^2\Psi_L, \cH_0 \cP^2 (\cH_0 + \Pi_L \cU_{\Psi_L}\Pi_L)\Psi_L}
    \nn 
    \\
    & = 2\Im \pscal{\Psi_L, \cH_0 \cU_{\Psi_L} \Psi_L}
		+ 2\Im \pscal{\cP^2 \Psi_L, \cH_0 \cP^2   \cU_{\Psi_L}\Psi_L}
        .
    \label{eqn.partial.t.E2.first.formula}
\end{align}
We state the more general bound on $\partial_t E_2(\Psi_L)$ as follows.

\begin{lemma}[Regularized energy propagation]
\label{lem.energy.general}
Let $V_{ee}$ satisfy \cref{assumption:Vee} and 
denote $|p|_M = \min\{|p|, M\}$ for any $M > 0$ and vector $p\in \R^D$. 
Then, for any real-valued $V\in H^2$ and $\Psi \in L^2_a$, 
\begin{multline*}
    \abs{2 \Im \pscal{\Psi, \cH_0 \cV \Psi} + 2 \Im \pscal{|\cP|_M^2 \Psi, \cH_0 |\cP|_M^2 \cV \Psi} - \int V \left(1 + |\nabla|_M^4 \right) \nabla \cdot j_\Psi}
    \\
    \leq C \norm{V}_{H^2} 
    \left(
    \norm{\pscal{|\cP|_M}^2\Psi}_{H^1}^2
    + \norm{\pscal{\cP}^2\Psi}_2^2
    \right). 
\end{multline*}
\end{lemma}

We give the proof below. With this, we now give the

\begin{proof}[Proof of \cref{lemma:energy}]
To prove \eqref{eq:bound_E_2} we use a Grönwall estimate. From \eqref{eqn.partial.t.E2.first.formula} and taking $M\to \infty$ in \cref{lem.energy.general} (note that all terms converge in this limit by \cref{lem:bilinear} and the equivalence of all norms from \cref{lemma:regularized}) we get 
\begin{equation*}
    \partial_t E_2(\Psi_L) \leq \int U_{\Psi_L} (1 + (-\Delta)^2) \nabla \cdot j_{\Psi_L} + C \norm{U_{\Psi_L}}_{H^2} E_2(\Psi_L). 
\end{equation*}
Recalling that $U_{\Psi_L} = V_{\Psi_L} - \partial_t \rho - \nabla \cdot j_{\Psi_L}$, integrating by parts twice, and bounding $V_{\Psi_L}$ using \cref{lemma:VPsi}, we thus find 
\begin{align}
    \partial_t E_2(\Psi_L) 
    & \leq - \norm{\nabla\cdot j_{\Psi_L}}_{H^2}^2
    + \norm{V_{\Psi_L} - \partial_t \rho}_{H^2} \norm{\nabla\cdot j_{\Psi_L}}_{H^2} 
    \nn \\ & \qquad 
    + C \norm{V_{\Psi_L} - \partial_t \rho - \nabla \cdot j_{\Psi_L}}_{H^2} E_2(\Psi_L)
    \nn \\ & \leq 
    - \norm{\nabla \cdot j_{\Psi_L}}_{H^2}^2 
    + C (E_2(\Psi_L) + \norm{\rho}_\cD) \norm{\nabla \cdot j_{\Psi_L}}_{H^2} + C (E_2(\Psi_L) + \norm{\rho}_\cD)^2. 
    \label{eqn.bdd.partial.t.E2}
\end{align}
By a completion of the square we can absorb the linear term in $\nabla \cdot j_{\Psi_L}$ in the quadratic one and finally drop it for an upper bound using its sign. We conclude the bound
\begin{equation*}
    \partial_t E_2(\Psi_L) \leq C E_2(\Psi_L)^2 + C \norm{\rho}_\cD^2,
\end{equation*}
where we note that, crucially, the constant does not depend on $L$. By Grönwall's inequality, we have $E_2(\Psi_L(t)) \leq z(t)$ with $\partial_t z = C z^2 + C \norm{\rho}_\cD^2$ and $z(0) = E_2(\Psi_L(0)) \leq E_2(\Pi_L\Psi_0)/\norm{\Pi_L\Psi_0}_2^2 \leq 4 E_2(\Psi_0)$. 
This ODE blows up at some finite time $T^*$ but stays uniformly bounded on $[0,T^*/2]$, say. Thus, $E_2(\Psi_L(t))$ is uniformly bounded for all $t\leq T' = \min\{T, T^*/2\}$ proving \eqref{eq:bound_E_2}.

Next, to prove \eqref{eq:bound_j}, we have from \eqref{eqn.bdd.partial.t.E2}, using the bound \eqref{eq:bound_E_2} and absorbing again the linear term in the quadratic and constant one 
\begin{equation*}
    \frac{1}{2}\norm{\nabla \cdot j_{\Psi_L}}_{H^2}^2 
    - C + \partial_t E_2(\Psi_L) 
    \leq 0,
\end{equation*}
where the constant $C$ depends on $\norm{\rho}_\cD$, but not on $L$ or $t$. 
Integrating in time, the bound in \eqref{eq:bound_j} follows.

Next, \eqref{eq:bound_U} follows immediately since we may bound 
\begin{equation*}
\norm{U_{\Psi_L(t)}(t)}_{H^2} 
\leq \norm{V_{\Psi_L(t)}(t)}_{H^2}
 + \norm{\partial_t \rho}_{H^2}
 +\norm{\nabla\cdot j_{\Psi_L(t)}}_{H^2}.
\end{equation*}
To prove next \eqref{eq:bound_H^2}, we compute, using \cref{lemma:embeddings},  
\begin{equation*}
\begin{split}
    \partial_t \la \Psi_L, \cH_0^2 \Psi_L\ra
    &= 2\Im \pscal{\Psi_L, \cH_0 \left[\cH_0, \cU_{\Psi_L} \right] \Psi_L}
    \leq C \norm{\cH_0 \Psi_L}^2_2 \norm{U_{\Psi_L(t)}(t)}_{H^2}. 
\end{split}
\end{equation*}
Grönwall's inequality together with $
\int_0^{T'} \rd t  \, \norm{U_{\Psi_L(t)}(t)}_{H^2} \leq C$, which follows directly from \eqref{eq:bound_U}, and $\norm{\Psi_L(0)}_{H^2} \leq C \norm{\Psi_0}_{H^2}$ then imply \eqref{eq:bound_H^2}.

Finally, this bound in turn implies \eqref{eq:bound_partial_t_Psi} since
	$$
	\norm{\partial_t \Psi_L}_2 
	\leq \norm{\cH_0 \Psi_L}_2
	+ \norm{\cU_{\Psi_L}\Psi_L}_2 
	\leq C \norm{\Psi_L}_{H^2}
	\left(1+\norm{U_{\Psi_L}}_2\right)
	$$
and $\norm{U_{\Psi_L}}_2 \leq C \norm{V_{\Psi_L}}_2
+\norm{\partial_t\rho}_2
+\norm{\nabla\cdot j_{\Psi_L}}_2
\leq C(E_2(\Psi_L)+ \norm{\rho}_{\cD})$ by \cref{lemma:VPsi,lem:bilinear}. 
\end{proof}

Finally, we give the 

\begin{proof}[Proof of \cref{lem.energy.general}]
Note first that $|\cP|_M$ commutes with $\cH_0$. We split $\cH_0 = \cT + \cV_{ee}$ according to the kinetic and interaction parts. Write, since $V_{ee}$ and $V$ are real, 
\begin{align*}
    & 2 \Im \pscal{\Psi, \cH_0 \cV \Psi} + 2 \Im \pscal{|\cP|_M^2 \Psi, \cH_0 |\cP|_M^2 \cV \Psi} 
    \\ & \quad 
    = 2 \Im \pscal{\cT\left(1 + |\cP|_M^4\right) \Psi, \cV \Psi}
    + 2 \Im \pscal{\cV_{ee} |\cP|_M^2 \Psi, \left[|\cP|_M^2,  \cV\right] \Psi}
    = : A_1 + A_2. 
\end{align*}
We compute first $A_2$. By \cref{assumption:Vee} (applied in one variable) we have 
\begin{equation*}
    \norm{\cV_{ee} |\cP|_M^2 \Psi}_2 \leq C \norm{|\cP|_M^2 \Psi}_{H^1} \leq C \pscal{|\cP|_M^2 \Psi, \cH_0 |\cP|_M^2 \Psi}^{1/2}. 
\end{equation*}
For the second factor, we may replace $\cV$ by $NV(x_1)$ by symmetry and bound the square sum of the Fourier coefficients of $[|\cP|_M^2,  V(x_1) ] \Psi$. These Fourier coefficients are $\sum_{p\in 2\pi \Z^D} \hat V(p) (|P|_M^2 - |P-p|_M^2) \hat \Psi(\vec p - p)$, where we denote $\vec p = (p_1,\ldots,p_N)$, $\vec p - p = (p_1-p, p_2,\ldots,p_N)$, and $P = \sum_{j=1}^N p_j$. 
Noting that $|\cdot|_M$ satisfies the triangle inequality we have (using the reverse triangle inequality)
\begin{multline*}
    \abs{|P|_M^2 - |P-p|_M^2} 
    \leq |p|_M ( |P|_M + |P-p|_M) \leq C |p|_M^2 + C |P-p|_M^2 
    \\
    \leq C \pscal{p}^2 \pscal{P-p}^2 \left(\pscal{p}^{-2} + \pscal{P-p}^{-2}\right). 
\end{multline*}
Computing the square sum of the Fourier coefficients and bounding the sums as in \cref{lemma:embeddings} we conclude that $\norm{\left[|\cP|_M^2,  \cV\right] \Psi}_2 \leq C \norm{V}_{H^2} \norm{\pscal{\cP}^2 \Psi}_{2}$ and thus that 
\begin{equation*}
    |A_2| \leq C \norm{V}_{H^2} \pscal{|\cP|_M^2 \Psi, \cH_0 |\cP|_M^2 \Psi}^{1/2} \norm{\pscal{\cP}^2 \Psi}_{2}. 
\end{equation*}

Next, consider $A_1$. Replacing $\cV$ by $NV(x_1)$ by symmetry and writing in Fourier variables, we have 
\begin{align*}
    A_1 & = 2 N \Im \sum_{\vec p, p} |\vec p|^2 \left(1 + |P|_M^4\right)\overline{\hat \Psi(\vec p)} \hat V(p) \hat \Psi(\vec p - p). 
\end{align*}
Using $\Im z = \frac{-i}{2} (z - \overline{z})$, that $\overline{\hat V(p)} = \hat V(-p)$ by reality of $V$, and changing variables $(\vec p, p) \to (\vec p - p, -p)$ we find
\begin{align*}
    A_1 & = - i N \sum_{\vec p, p}  \left[|\vec p|^2\left(1 + |P|_M^4\right) - |\vec p - p|^2 \left(1 + |P-p|_M^4\right) \right]\overline{\hat \Psi(\vec p)} \hat V(p) \hat \Psi(\vec p - p).
\end{align*}
Next, using that 
$\nabla\cdot j_\Psi = 2 \Im \rho_{\Psi, \Delta_{x_1} \Psi} 
= - i (\rho_{\Psi, \Delta_{x_1}\Psi} - \rho_{\Delta_{x_1}\Psi, \Psi})$, 
and $\hat \rho_{\Psi_1,\Psi_2}(p) = N \sum_{\vec p} \overline{\hat\Psi_1(\vec p)} \hat \Psi_2(\vec p + p)$, 
we have
\begin{align*}
    A_{11} & := \int V \left(1 + |\nabla|_M^4 \right) \nabla \cdot j_\Psi 
    =  \sum_{p\in 2\pi \Z^D} \overline{\hat V(-p)} \left(1 + |p|_M^4 \right) \hat{\nabla\cdot j_\Psi}(-p)
    \\ & \hphantom{:}= 
    -iN \sum_{\vec p, p}\hat V(p) \left(1 + |p|_M^4 \right) \left(-|p_1-p|^2 + |p_1|^2\right) \overline{\hat \Psi(\vec p)}  \hat \Psi(\vec p -p)
    \\ & \hphantom{:}= 
    -iN \sum_{\vec p, p} \left(1 + |p|_M^4 \right) \left(|\vec p|^2 - |\vec p-p|^2\right) \overline{\hat \Psi(\vec p)}  \hat V(p)  \hat \Psi(\vec p -p). 
\end{align*}
Thus, we need to bound $A_{12} := A_1 - A_{11} $ given by 
\begin{align*}
    A_{12} = - i N \sum_{\vec p, p} \left[|\vec p|^2 \left(|P|_M^4 - |p|_M^4\right) - |\vec p - p|^2\left(|P-p|_M^4 - |p|_M^4\right)\right] \overline{\hat \Psi(\vec p)}  \hat V(p)  \hat \Psi(\vec p -p). 
\end{align*}
Let $\Xi = \left[|\vec p|^2 \left(|P|_M^4 - |p|_M^4\right) - |\vec p - p|^2\left(|P-p|_M^4 - |p|_M^4\right)\right]$. We claim that  
\begin{equation}
    |\Xi| \leq C \pscal{p}^2 \pscal{\vec p} \pscal{|P|_M}^2 \pscal{\vec p -p} \pscal{|P-p|_M}^2 \left(\frac{1}{\pscal{p}^2} + \frac{1}{\pscal{P}^2} + \frac{1}{\pscal{P-p}^2}\right). 
\label{eqn.bdd.Xi.claimed}
\end{equation}
With this bound and Cauchy--Schwarz we find 
\begin{equation*}
    |A_{12}| \leq C \norm{V}_{H^2} \norm{\pscal{|\cP|_M}^2 \Psi}_{H^1}^2
    \leq C \norm{V}_{H^2} \left( \pscal{\Psi, \cH_0\Psi} + \pscal{|\cP|_M^2 \Psi, \cH_0 |\cP|_M^2 \Psi} \right)
\end{equation*}
as desired. It remains to prove \eqref{eqn.bdd.Xi.claimed}.

First note that the involution $(\vec p, p) \to (\vec p - p, -p)$ changes the sign of $\Xi$ and leaves the right-hand-side of \eqref{eqn.bdd.Xi.claimed} invariant. Thus, we may without loss of generality assume $|P-p|\leq |P|$. It follows that 
\begin{equation}\label{eqn.bdd.by.P}
    |p| \leq C |P|, \qquad 
    |p|_M \leq C |P|_M,  \qquad 
    |\vec p - p| \leq C |\vec p|. 
\end{equation}
Indeed, the first and second inequalities follow by $|p| \leq |p-P| + |P| \leq 2 |P|$. 
To prove the last, we note that $|P|\leq \sqrt{N}|\vec p|$. 
Then, $|\vec p - p| \leq |\vec p| + |p| \leq |\vec p| + C |P| \leq C |\vec p|$ as claimed.

Next, write 
\begin{equation*}
    \Xi = \left(|\vec p|^2 - |\vec p - p|^2\right)\left(|P|_M^4 - |p|_M^4\right)
        + |\vec p - p|^2\left(|P|_M^4 - |P-p|_M^4\right). 
\end{equation*}
We bound the two terms separately. We have by the (reverse) triangle inequality and \eqref{eqn.bdd.by.P}
\begin{equation*}
    \abs{|\vec p|^2 - |\vec p - p|^2} \leq |p| (|\vec p| + |\vec p - p|) \leq C |p| |\vec p|
\end{equation*}
and
\begin{equation*}
    \abs{|P|_M^4 - |p|_M^4} \leq |P-p|_M (|P|_M^3 + |P|_M^2|p|_M + |P|_M|p|_M^2 + |p|_M^3) \leq C |P-p|_M |P|_M^3
\end{equation*}
With  a similar bound for the second term in $\Xi$ we find 
\begin{equation}\label{eqn.bdd.Xi}
    |\Xi| 
    \leq C |P|_M^3 |\vec p| \left(|p||P-p|_M  + |\vec p - p| |p|_M\right). 
\end{equation}
We now split into two cases: 

\underline{If $|p| \leq |P-p|$:} Then $|p|_M \leq |P-p|_M$ and thus also $|P|_M \leq 2 |P-p|_M$. Moreover, $|p|\leq |P-p| \leq \sqrt{N}|\vec p -p|$. 
Thus, from \eqref{eqn.bdd.Xi}, 
\begin{equation*}
    |\Xi| \leq C |P|_M^2 |P-p|_M^2 |\vec p| |\vec p - p|
\end{equation*}
which satisfies \eqref{eqn.bdd.Xi.claimed}. 

\underline{If $|p| > |P-p|$:} It follows that $|P|_M \leq 2 |p|_M$. Moreover, we note that $\pscal{P-p}\leq C \pscal{\vec p - p}$. We multiply and divide $\Xi$ by $\pscal{P-p}^2$ and use that 
\begin{equation*}
    \textnormal{If $p,q$ satisfy $|q|\leq |p|$, then $|p|_M\pscal{q} \leq \pscal{p} \pscal{|q|_M}$}
\end{equation*}
for $q=P-p$. (This implication is easily seen.)
Then, from \eqref{eqn.bdd.Xi},
\begin{align*}
    \pscal{P-p}^2 |\Xi| 
    & \leq C |P|_M^2 |\vec p| \left( |p|_M |p| |P-p|_M \pscal{P-p}^2 
    + |p|_M^2 |\vec p - p| \pscal{P-p}^2 
    \right)
    \\ & \leq C |P|_M^2 |\vec p| \pscal{|P-p|_M}^2 \pscal{\vec p -p} \pscal{p}^2. 
\end{align*}
That is, \eqref{eqn.bdd.Xi.claimed} is satisfied. This concludes the proof. 
\end{proof}

\subsection{Removing the regularization}
Using the energy bounds above, we can now give the proof that the regularized solutions converge when removing the regularization.

\begin{lemma}[Convergence of regularized sequence]\label{lemma:convergence_regularized}
Let $V_{ee},\Psi_0, \rho, T$ be as in \cref{thm.main}.
	Let $\Psi_L(t)$ be the solutions to \eqref{eq:SEnonlinear_regularized} from \cref{lemma:regularized}.
	Then there is a $0<T'\leq T$ and 
    $\Psi_\infty \in C^0([0,T'];H^2) \cap L^\infty([0,T'];H^1_2) \cap C^1([0,T'];L^2)$ 
    such that along a subsequence $\Psi_L\to \Psi_\infty$ in
    $C^0([0,T'];H^1_1)$.
	The limit wavefunction $\Psi_\infty$ solves \eqref{eqn.nonlinear.SE.U} in the Duhamel form: 
	\begin{equation}\label{eqn.duhamel.limit}
	\Psi_\infty (t) 
	=e^{-i \cH_0 t} \Psi_0 
    -i 
    \int_0^t \rd s \,  e^{-i \cH_0 (t-s) }  \cU_{\Psi_\infty(s)}(s) \Psi_\infty(s).
\end{equation}
Moreover, 
$V_{\Psi_\infty} \in L^\infty([0,T']; H^2)$
and $U_{\Psi_\infty}\in L^2([0,T'];H^2) \cap L^\infty([0,T'];H^1) $. 
\end{lemma}

\begin{proof}
By \cref{lemma:regularized,lemma:energy}, the sequence $\Psi_L \in C^0([0,T'];H^2\cap H_2^1)$ is uniformly bounded 
and equicontinuous as a sequence in $C^0([0,T']; L^2)$.

The Arzelà--Ascoli theorem thus implies that there is $\Psi_\infty \in C^0([0,T']; L^2)$ such that along a subsequence (we omit this from the notation) $\Psi_L \to \Psi_\infty$ uniformly in $[0,T']$. 
By lower semicontinuity of the norm we have $\Psi_\infty \in L^\infty([0,T']; H^2 \cap H^1_2)$
and by interpolation we find 
$\Psi_L(t) \to \Psi_\infty(t)$ in $H^1_1$ uniformly in $t$.

We consider next the nonlinear potentials associated with $\Psi_L$ and $\Psi_\infty$.
By \cref{lemma:VPsi} we find $V_{\Psi_\infty} \in L^\infty([0,T']; H^2)$
and $V_{\Psi_L} \to V_{\Psi_\infty}$ in $C^0([0,T'];L^2)$. 
Next, by assumption, $\partial_t \rho \in C^0([0,T']; H^2)$ and furthermore, by \cref{lem:bilinear} and the convergence of $\Psi_L$ in $H^1_1$, we find 
$\nabla\cdot j_{\Psi_L} \to \nabla\cdot j_{\Psi_\infty}$ in $C^0([0,T']; L^2)$ and 
$\nabla\cdot j_{\Psi_\infty}\in L^\infty([0,T']; H^1)$. 
In particular, also $U_{\Psi_\infty} \in C^0([0,T'];L^2) \cap  L^\infty([0,T'];H^1)$.
Now, \eqref{eq:bound_U} in \cref{lemma:energy} shows that $U_{\Psi_L}$ is bounded as a sequence in $L^2([0,T'];H^2)$ and we may thus extract a weakly convergent subsequence by Banach--Alaoglu. By weak lower semi-continuity of the norm we thus have $U_{\Psi_\infty} \in L^2([0,T'];H^2)$.

Next, to prove that $\Psi_\infty$ satisfies \eqref{eqn.duhamel.limit}, we pass to the limit weakly in \eqref{eq:SEnonlinear_regularized}. First, from the uniform-in-time convergence of $V_{\Psi_L}$ and $\nabla\cdot j_{\Psi_L}$ in $L^2$ and of $\Psi_L$ in $H^1_1$ we find by \cref{lemma:embeddings} that $\cU_{\Psi_L}\Psi_L$ converges in $L^2$. 
Testing \eqref{eq:SEnonlinear_regularized} against any $H^{1}([0,T'];H^2)$-function and noting that $\Pi_L\to \mathbbm{1}$ strongly on $H^2$, we find $i \partial_t \Psi_\infty = \cH_0 \Psi_\infty +  \cU_{\Psi_\infty}\Psi_\infty$, where the time-derivative is to be interpreted weakly. Writing this in Duhamel form, we get \eqref{eqn.duhamel.limit} as claimed.

Next, to prove that $\Psi_\infty \in C^0([0,T'];H^2)$ we use  \eqref{eqn.duhamel.limit}. First, we bound by \cref{lemma:embeddings}
$\norm{\cU_{\Psi_\infty}\Psi_\infty}_{H^2} \leq C \norm{U_{\Psi_\infty}}_{H^2} \norm{\Psi_\infty}_{H^2} \leq C \norm{U_{\Psi_\infty}}_{H^2}$. 
Next, by $U_{\Psi_\infty} \in L^2([0,T'];H^2)\subset L^1([0,T'];H^2)$ we have 
$\int_0^t \rd s \, \norm{\cU_{\Psi_\infty} \Psi_\infty}_{H^2} \leq C$
uniformly in $t\leq T'$. In particular, the integrand is an $L^1$-function in time and the integral in \eqref{eqn.duhamel.limit} is thus a continuous function of time in the $H^2$ topology. We conclude that $\Psi_\infty\in C^0([0,T'];H^2)$ as claimed.

Finally, to see that $\Psi_\infty \in C^1([0,T'];L^2)$ we again use \eqref{eqn.duhamel.limit}. We prove that $\cU_{\Psi_\infty}\Psi_\infty$ is continuous in time in the $L^2$-topology. 
This follows from \cref{lemma:embeddings}, the continuity of $U_{\Psi_\infty}$ in $L^2$, and that of $\Psi_\infty$ in $H^2$, hence also in $H^1_1$. We conclude that $\Psi_\infty \in C^1([0,T'];L^2)$ as claimed.
\end{proof}

\subsection{Uniqueness of solutions}
Next, we prove the claimed uniqueness of solutions. To do this, it will be useful to show first that all solutions to \eqref{eqn.nonlinear.SE.U} are in fact also solutions to \eqref{eq:SEnonlinearV}.

\begin{lemma}[Equality of densities and potentials]
\label{lemma:density_reproduced}
Let $V_{ee},\Psi_0, \rho, T$ be as in \cref{thm.main} and
let $\Psi \in C^0([0,T']; H^2) \cap L^\infty([0,T'];H^1_2) \cap C^1([0,T']; L^2)$ be a solution to \eqref{eqn.nonlinear.SE.U}. 
Then, $\rho_{\Psi}=\rho$. 
In particular, $ U_{\Psi} =  V_{\Psi} $ and thus, $\Psi$ is a solution to \eqref{eq:SEnonlinearV} as well. 
\end{lemma}

\begin{proof}
To show that $\rho_{\Psi}=\rho$ we denote the difference $h = \rho_{\Psi} - \rho$ and note that by the initial compatibility assumptions of \cref{thm.main}, we have $h(0) = \partial_t h(0) = 0$. 
By the continuity equation, the first derivative is $\partial_t h = - \nabla\cdot j_{\Psi} -\partial_t\rho$. Thus, recalling \eqref{eq:UPsi}, we have $U_{\Psi} = V_{\Psi} + \partial_t h$.
Next, using the force-balance equation (compare \cite[Prop.~5.1]{LauLewOld-V-26_ppt}) the second derivative is 
    $$
  \partial_t^2\rho_{\Psi}  
  = - \Delta^2\rho_{\Psi} - 2K_{\rho_{\Psi}}U_{\Psi} + Q_{\Psi}. 
$$
(More precisely, this is an equality between distributions on $\Lambda$. It is proved exactly as \cite[Prop.~5.1]{LauLewOld-V-26_ppt}.)
Inserting $U_{\Psi} = V_{\Psi} + \partial_t h$ and recalling the definition of $V_{\Psi}$ in \eqref{eq:VPsi} we obtain
\begin{align*}
    \partial_t^2 h 
    & 
    = -\Delta^2 h
  +2\nabla \cdot (h\nabla V_{\Psi})
  +2 \nabla \cdot (\rho_{\Psi} \nabla\partial_t h). 
\end{align*}
To show that $h=0$ is the unique solution to this equation we define the energy $E(t) = \la h, (1+\Delta^2) h\ra + \la \partial_t h, \partial_t h\ra$ and use a Grönwall argument.
Note that we have $h\in L^\infty([0,T'];H^3)$, $\partial_t h \in L^\infty([0,T'];H^1)$, and $\partial^2_t h \in L^\infty([0,T'];H^{-1})$ by \cref{lem:bilinear} and $V_\Psi\in L^\infty([0,T'];H^2)$. 
Thus, by the Lions--Magenes theorem \cite[Thm.~5.9.3]{Evans-22} we can compute
\begin{equation*}
\begin{split}
  \partial_t E(t)
  &= 2\Re \la \partial_t h, (1+\Delta^2) h\ra
   + 2\Re \la \partial_t h, \partial_t^2 h\ra\\
     &= 2\Re \la \partial_t h, h\ra
   + 4\Re \la \partial_t h, \nabla \cdot (h\nabla V_{\Psi}) \ra
   + 4 \la \partial_t h, \nabla \cdot (\rho_{\Psi}\nabla\partial_t h)\ra\\
     &\leq 2 \norm{\partial_t h}_2 \norm{h}_2
   + C \norm{\partial_t h}_2
    \norm{h}_{H^2}\norm{V_{\Psi}}_{H^2}
    \leq C(1+\norm{V_{\Psi}}_{H^2}) E(t).
\end{split}
\end{equation*}
Here we dropped the last term for an upper bound, since $\nabla \cdot(\rho_{\Psi} \nabla \cdot) = - K_{\rho_{\Psi}} \leq 0$. 
Using $\norm{V_{\Psi}}_{H^2} \leq C E_2(\Psi) + C \norm{\rho}_\cD < C$ by \cref{lemma:VPsi}, we conclude $E\equiv 0$ and hence $h\equiv 0$. Thus $\rho_{\Psi} = \rho$. 
Recalling the formula $U_{\Psi}= V_{\Psi} + \partial_t h$ we conclude the equality of the nonlinear potentials.  
\end{proof}

With this, we can prove the claimed uniqueness. 

\begin{lemma}[Uniqueness]\label{lemma:uniqueness_nonlinear}
Let $V_{ee},\Psi_0, \rho, T$ be as in \cref{thm.main} and let $\Psi,\tilde\Psi \in C^0([0,T'];H^2) \cap L^\infty([0,T'];H^1_2) \cap C^1([0,T'];L^2)$ be solutions to \eqref{eqn.nonlinear.SE.U}. Then $\Psi = \tilde\Psi$. 
\end{lemma}

\begin{proof}
Denote $\delta = \Psi - \tilde\Psi$. We show that $\norm{\delta}_{H^1}^2=E_0(\delta)=0$ using a Grönwall argument. First, clearly $\delta(0)=0$.
To compute the derivative $\partial_t E_0(\delta)$, we first note that, by \cref{lemma:density_reproduced}, both $\Psi$ and $\tilde\Psi$ solve \eqref{eq:SEnonlinearV}. Thus,   
\begin{align}
    \partial_t E_0(\delta) 
    & = 2\Im \pscal{\delta, \cH_0 (\cV_{\Psi} \Psi - \cV_{\tilde\Psi}\tilde\Psi)}
    = 2\Im \pscal{\delta, \cH_0 \cV_{\Psi} \delta} + 2 \Im \pscal{\delta, \cH_0 (\cV_\Psi - \cV_{\tilde\Psi})\tilde\Psi}
    .
    \label{eqn.compute.partial.t.E0}
\end{align}
For the first term we have by \cref{lemma:embeddings,lemma:VPsi}
\begin{align*}
    \Im \pscal{\delta, \cH_0 \cV_{\Psi} \delta} 
    & = \sum_{j,k=1}^N \Im \pscal{\nabla_{x_j} \delta, \nabla_{x_j} V_\Psi(x_k) \delta} 
    = \sum_{j=1}^N \Im \pscal{\nabla_{x_j} \delta, (\nabla V_\Psi)(x_j) \delta} 
    \\ & 
    \leq C \norm{\delta}_{H^1} \norm{V_\Psi}_{H^2} \norm{\pscal{\cP}\delta}_2
    \leq C E_0(\delta). 
\end{align*}
For the second term in \eqref{eqn.compute.partial.t.E0} we have  
\begin{align}
    \Im \pscal{\delta, \cH_0 (\cV_\Psi - \cV_{\tilde\Psi})\tilde\Psi} 
    & = 
    \sum_{j,k=1}^N \Im \pscal{\nabla_{x_k}\delta, \nabla_{x_k} (V_\Psi(x_j) - V_{\tilde\Psi}(x_j))\tilde\Psi} 
    \nn
    \\* & \quad 
    + \sum_{j,k,l=1 : k < l}^N \Im \pscal{ V_{ee}(x_k - x_l) \delta, (V_\Psi(x_j) - V_{\tilde\Psi}(x_j))\tilde\Psi}
    \nn \\ & 
    = : A_1 + A_2
    \label{eqn.decompose.delta.HVPsi.second}
\end{align}
The second term is bounded by \cref{assumption:Vee} (applied in just one coordinate $x_k$ with all others fixed) and \cref{lemma:embeddings,lemma:VPsi} as 
\begin{multline*}
    |A_2|
    \leq C \norm{\delta}_{H^1} \norm{V_\Psi - V_{\tilde \Psi}}_2 \snorm{\pscal{\cP}^2\tilde\Psi}_{2}
    \leq C \norm{\delta}_{H^1}^2 \left(\norm{\Psi}_{H^1_2} + \snorm{\tilde\Psi}_{H^1_2}\right)^2 \leq C E_0(\delta). 
\end{multline*}
Next, we write for the first term in \eqref{eqn.decompose.delta.HVPsi.second}
\begin{multline*}
    A_1
    = \sum_{j,k=1}^N \Im \pscal{\nabla_{x_j}\delta,  (V_\Psi(x_k) - V_{\tilde\Psi}(x_k))\nabla_{x_j}\tilde\Psi} 
    \\ 
    + \sum_{j=1}^N \Im \pscal{\nabla_{x_j}\delta, (\nabla V_\Psi(x_j) - \nabla V_{\tilde\Psi}(x_j))\tilde\Psi} 
    = : A_{11} + A_{12}
\end{multline*}
The first term here is bounded as $A_2$ above using that $\snorm{\pscal{\cP}^2\nabla_{x_j}\tilde\Psi}_2 \leq C\snorm{\tilde\Psi}_{H^1_2}$. For the second we write it as 
\begin{align*}
A_{12}
& = \sum_{\alpha=1}^D \int \rd x \, \partial^\alpha (V_\Psi - V_{\tilde\Psi}) \Im \rho_{\partial^\alpha \delta, \tilde\Psi} 
= - \int \rd x \, (V_\Psi - V_{\tilde\Psi}) \sum_{\alpha=1}^D \partial^\alpha \Im \rho_{\partial^\alpha \delta, \tilde\Psi} 
.
\end{align*}
Next, we note that both $\Psi$ and $\tilde\Psi$ satisfy the continuity equation and have $\rho_\Psi = \rho_{\tilde\Psi}$ by \cref{lemma:density_reproduced}, so 
\begin{equation*}
    0 = \nabla \cdot j_\Psi - \nabla \cdot j_{\tilde\Psi}
    = 2 \sum_{\alpha=1}^D \partial^\alpha 
    \Im \left( \rho_{\Psi, \partial^\alpha \Psi}  -\rho_{\tilde\Psi, \partial^\alpha \tilde\Psi} \right)
    = 2 \sum_{\alpha=1}^D \partial^\alpha 
    \Im \left( \rho_{\delta, \partial^\alpha \Psi}  - \rho_{\partial^\alpha\delta, \tilde\Psi} \right). 
\end{equation*}
Thus, we have the bound, using \cref{lem:bilinear,lemma:VPsi},  
\begin{align*}
    |A_{12}| & \leq C \norm{V_\Psi - V_{\tilde\Psi}}_2 \norm{\rho_{\delta, \nabla \Psi}}_{H^1} \leq C \norm{\delta}_{H^1}^2 \snorm{\Psi + \tilde\Psi}_{H_2^1} \norm{\Psi}_{H^1_2} \leq C E_0(\delta). 
\end{align*}
In conclusion, we have thus proved $\partial_t E_0(\delta) \leq C E_0(\delta)$. By Grönwall's inequality and the fact that $E_0(\delta(0)) = 0$ we conclude that $\delta=0$ as claimed. 
\end{proof}

\subsection{Improved regularity}
Finally, we show that the solution $\Psi_\infty$ is in fact continuous in time in the $H^1_2$-topology. For use in the proof of \cref{thm.main} below, we formulate a slightly more general result.  

\begin{lemma}[Regularity of density-reproducing states]
\label{lem.direct}
Let $V_{ee},\Psi_0, \rho, T$ be as in \cref{thm.main},
let $V\in L^\infty([0,T'];H^2)$, 
and let $\Psi \in C^0([0,T'];H^2) \cap C^1([0,T'];L^2)$ be a solution to  \eqref{eq:Schrodinger} $i \partial_t \Psi = \cH_V \Psi$ with $\Psi(t=0)=\Psi_0$. 
Assume that $\rho_{\Psi} = \rho$. Then, $\Psi \in C^0([0,T'];H^1_2)$. 
\end{lemma}
\begin{proof} 
This essentially follows from the regularized energy propagation in \cref{lem.energy.general} and removing the regularization as in \cite[Lem.~4.4]{LauLewOld-V-26_ppt}. More precisely, defining $E_2^M(\Psi) = \pscal{\Psi, \cH_0\Psi} + \pscal{|\cP|_M^2 \Psi, \cH_0 |\cP|_M^2 \Psi}$, we have, by a similar computation as in \eqref{eqn.partial.t.E2.first.formula}, (noting that $E_2^M(\Psi)$ is $C^1$ in time since $|\cP|_M$ is a bounded operator)
\begin{equation*}
    \partial_t E_2^M(\Psi) = 2 \Im \pscal{\Psi, \cH_0 \cV \Psi} + 2 \Im \pscal{|\cP|_M^2 \Psi, \cH_0 |\cP|_M^2 \cV \Psi}. 
\end{equation*}
Using that the density is reproduced we have the continuity equation $\nabla \cdot j_\Psi = - \partial_t \rho$ and thus, by integration by parts, 
\begin{equation*}
    \abs{\int V \left(1 + |\nabla|_M^4\right) \nabla \cdot j_\Psi \rd x} \leq \norm{V}_{H^2} \norm{\partial_t \rho}_{H^2}.
\end{equation*}
Thus, by \cref{lem.energy.general} and bounding $\pscal{\cP}^2 \leq C (1+\cT)$, 
\begin{equation} \label{eqn.bdd.partial.t.E2.regularized}
    |\partial_t E_2^M(\Psi)| \leq C \norm{V}_{H^2} \left(\norm{\rho}_\cD + E_2^M(\Psi) + \norm{\Psi}_{H^2}^2\right). 
\end{equation}
By first Grönwall's inequality and then taking $M\to \infty$ we find $\Psi \in L^\infty([0,T'];H^1_2)$. Moreover, since $\Psi$ is continuous in time in the $L^2$-topology we find that $\Psi$ is weakly continuous in time in the $H^1_2$-topology. 
Next, taking $M\to \infty$ in \eqref{eqn.bdd.partial.t.E2.regularized} we see that $\partial_t E_2(\Psi)$ is bounded and hence that $E_2(\Psi)$ is continuous in time. Now, $H^1_2$ is a Hilbert space, so weak convergence and convergence of the norms together give norm-convergence and we conclude that $\Psi\in C^0([0,T'];H^1_2)$ as claimed. 
\end{proof}

With this we can now prove 

\begin{lemma}[Upgraded continuity]
\label{lem.upgraded.continuity}
Let $\Psi_\infty$ denote the limiting solution to \eqref{eqn.nonlinear.SE.U} constructed in \cref{lemma:convergence_regularized}. Then, $\Psi_\infty \in C^0([0,T'];H^1_2)$ and 
$V_{\Psi_\infty} \in C^0([0,T']; H^2)$.
\end{lemma}

\begin{proof}
By \cref{lemma:convergence_regularized}, $\Psi_\infty \in C^0([0,T'];H^2)\cap L^\infty([0,T'];H^1_2)\cap C^1([0,T'];L^2)$ and $V_{\Psi_\infty}\in L^\infty([0,T'];H^2)$. Then by \cref{lemma:density_reproduced} we have $\rho_{\Psi_\infty} = \rho$ and thus that $\Psi_\infty$ solves \eqref{eq:Schrodinger} with $V=V_{\Psi_\infty}\in L^\infty([0,T'];H^2)$. 
Hence, by \cref{lem.direct}, we have $\Psi_\infty \in C^0([0,T'];H^1_2)$ as claimed. Finally, $V_{\Psi_\infty}\in C^0([0,T'];H^2)$ by \cref{lemma:VPsi}.  
\end{proof}

\subsection{Final steps in the proof of \texorpdfstring{\cref{thm.main}}{Theorem \ref{thm.main}}}
\label{sec.final.steps.main.thm}
Finally, we give the 

\begin{proof}[Proof of \cref{thm.main}]
We use the above lemmas studying the nonlinear equation \eqref{eqn.nonlinear.SE.U}.

\smallskip
\noindent
\underline{Existence:} 
Let $\Psi_\infty$ denote the solution to \eqref{eqn.nonlinear.SE.U} constructed in \cref{lemma:convergence_regularized,lem.upgraded.continuity}. We define the potential $V = V_{\Psi_\infty}$ and claim that this satisfies the desiderata. By construction $\int V = 0$ and by \cref{lem.upgraded.continuity}, $V\in C^0([0,T'];H^2)$ as desired. It remains to be shown that this potential correctly reproduces the density. Thus, consider the solution $\Psi$ to the Schrödinger equation \eqref{eq:Schrodinger} $i\partial_t \Psi = \cH_{V} \Psi$. 
By \cref{lemma:density_reproduced},  $V = V_{\Psi_\infty} = U_{\Psi_\infty}$, so $\Psi_\infty$ also solves $i\partial_t \Psi_\infty = \cH_V \Psi_\infty$. Thus, $\Psi = \Psi_\infty$ by uniqueness of solutions to the linear equation, see \cref{rmk.direct}.
Then, by \cref{lemma:density_reproduced} we have $\rho_\Psi = \rho_{\Psi_\infty} = \rho$ as desired.

\smallskip
\noindent
\underline{Uniqueness:}
Let $\tilde V\in  C^0([0,T']; H^2)$ with $\int \tilde V = 0$ be a potential with associated solution $\tilde \Psi \in C^0([0,T'];H^2)\cap C^1([0,T'];L^2)$ to \eqref{eq:Schrodinger} (recall \cref{rmk.direct}) such that $\rho_{\tilde\Psi} = \rho$. 
By \cref{lem.direct} we have $\tilde\Psi \in C^0([0,T'];H^2\cap H^1_2) \cap C^1([0,T'];L^2)$. 
Next, by the force-balance equation (which is well-founded as an equality in $L^2$ by the regularity of $\tilde\Psi$ combined with $\rho_{\tilde\Psi} = \rho$, see the proof of \cite[Prop.~5.1]{LauLewOld-V-26_ppt})
\begin{equation*}
    \partial_t^2 \rho = \partial_t^2 \rho_{\tilde\Psi}
    = -\Delta^2 \rho_{\tilde\Psi} - 2 K_{\rho_{\tilde\Psi}} \tilde V + Q_{\tilde \Psi}
    = - \Delta^2 \rho - 2 K_\rho \tilde V + Q_{\tilde \Psi}
\end{equation*}
we find $\tilde V = V_{\tilde\Psi} = U_{\tilde\Psi}$ (recalling $\int \tilde V = 0 = \int V_{\tilde\Psi}$ and the continuity equation) and thus by the uniqueness of solution of \eqref{eqn.nonlinear.SE.U} in \cref{lemma:uniqueness_nonlinear} we have $\tilde\Psi =  \Psi_\infty$ and thus $\tilde V = V_{\tilde \Psi} = V_{\Psi_\infty} = V$.
\end{proof}

\printbibliography

\end{document}